\documentclass[twocolumn, preprintnumbers, amsmath, amssymb]{revtex4}
\usepackage{graphicx}
\begin{document}
\preprint{KOBE-COSMO-16-02}
\title{Pulsar timing signal from ultralight axion in $f(R)$ theory}
\author{Arata Aoki and Jiro Soda}
\affiliation{Department of Physics, Kobe University, Kobe 657-8501, Japan}
\date{\today}

\begin{abstract}
An ultralight axion around $10^{-23}$ eV is known as a viable dark matter candidate.
A distinguished feature of such a dark matter is the oscillating pressure which produces the oscillation of the gravitational potential with frequency in the nano-Hz range.
Recently, Khmelnitsky and Rubakov pointed out that this time dependent potential induces the pulse arrival residual and could be observed by the Square Kilometre Array (SKA) experiment.
In this paper, we study the detectability of the oscillating pressure of the axion in the framework of $f(R)$ theory, and show that the amplitude of the gravitational potential can be enhanced or suppressed compared to that in Einstein's theory depending on the parameters of the $f(R)$ model and mass of the axion.
In particular, we investigate the Hu-Sawicki model and find the condition that the Hu-Sawicki model is excluded.
\end{abstract}

\maketitle

\section{Introduction}
The dark energy and dark matter problems are the most important unsolved issues in cosmology.
Although many models of dark energy and dark matter have been proposed, none of them are conclusive at the present time.
Phenomenologically, the cold dark matter with a cosmological constant ($\Lambda$CDM model) is currently the most successful cosmological model.
The most promising candidate of the cold dark matter (CDM) is supersymmetric particles, the so-called neutralino.
While CDM works quite well especially on large scales, it is known that there exists a problem on small scales.
In fact, this model predicts a cusp of dark matter halo profile and overabundance of dwarf galaxies, which are not consistent with observations.
Moreover, the LHC has not reported any signature of supersymmetry.
Given this situation, it is worth seeking another possibility, namely axion dark matter.

The axion, the pseudo-Nambu-Goldstone boson, is originally introduced by Peccei and Quinn to resolve the strong CP problem of QCD~\cite{770620}.
Nowadays, it is known that the string theory also predicts such scalar fields with a wide range of mass scales~\cite{060626}.
Note that we use the term ``axion" in more general meaning, e.g., axionlike particles and other ultralight scalar particles.
The axion with the mass around $10^{-23}~\text{eV}$ behaves as nonrelativistic matter on cosmological scales, and hence it can be regarded as a candidate of dark matter.
Furthermore, it is known that such an ultralight axion can resolve the cusp problem on subgalactic scales because of its wave nature~\cite{000807, 14:Schive}.
A peculiarity of the axion is the oscillating pressure in time with frequency at the twice of the axion mass, $2m$.
Therefore, in order to identify the axion dark matter, we should detect the effect of the oscillating pressure of the axion.
The period of the oscillation corresponds to about one year, and this time scale is much shorter than the cosmological time scale, i.e. $H_{0}^{-1} \sim 10^{10}~\text{years}$.
Hence, after averaging the oscillating pressure over the cosmological time scale, the axion behaves as pressureless dust on cosmological scales.
Thus, it might be difficult to distinguish the axion from other dark matter candidates by cosmological scale observations.
For this reason, we should pay attention to smaller scales to prove the existence of the axion dark matter.
From this point of view, it is pointed out by Khmelnitsky and Rubakov that the effect of oscillating pressure of the axion can be detected by pulsar timing array experiments~\cite{140212}.
The oscillating pressure induces the oscillation of the gravitational potential with frequency in the nano-Hz range.
This effect can be observed as a shift of the arrival time of the signal from the pulsar.

In the previous paper, the axion oscillation was studied in Einstein's theory.
However, since the main energy component of the universe is the dark energy, it might be necessary to consider this issue in the context of modified gravity.
The reason is as follows:
The simplest candidate of dark energy, i.e., the cosmological constant, has several problems, e.g. the fine-tuning problem and coincidence problem.
One possibility to resolve these issues is to consider unknown matter such as the quintessence.
However, there is no natural candidate of quintessence in particle physics.
Therefore, it is natural to assume that theory of gravity is different from Einstein's theory on cosmological scales.

In this paper, as a first step to this direction, we focus on the $f(R)$ theory, which is the simplest modified gravity.
We discuss the detectability of the oscillation of the gravitational potential induced by the time-oscillating pressure of the axion in this context.

This paper is organized as follows:
In Sec. II, we review the results obtained by Khmelnitsky and Rubakov in Einstein's theory.
In Sec. III, we formulate the procedure to determine the amplitude of the gravitational potential in the framework of $f(R)$ theory and discuss two specific models: $R^{2}$ model which can be solved exactly, and the Hu-Sawicki model which is known as a viable cosmological model.
The final section is devoted to conclusion.

\section{Pulsar Timing and Ultralight Axions in Einstein's Theory}
In this section, we review the results obtained by Khmelnitsky and Rubakov in Einstein's theory~\cite{140212}.

We consider the situation that the dark matter halo is composed out of a free ultralight axion field.
The trace of the Einstein equation gives
\begin{equation}
R = -T \ ,
\label{eq1}
\end{equation}
where $R$ is the Ricci scalar and $T$ is the trace of the energy-momentum tensor of the axion field.
We will use this equation to determine the gravitational potentials with given $T$.
Now let us consider both sides of this equation in turn.

On the scale of the dark matter halo, the expansion of the universe is completely negligible and the gravitational potentials can still be treated as perturbation.
Thus, we use the Newtonian gauge for the metric:
\begin{equation}
g_{\mu\nu} = \left( \begin{array}{cc}
-1 - 2\Psi & 0 \\ 0 & (1 - 2\Phi)\delta_{ij}
\end{array} \right).
\label{eq2}
\end{equation}
Note that this convention is different from that in \cite{140212}: it is $\Phi$ that affects the signal from the pulsar in this paper.
The Ricci scalar $R$ can be calculated from the metric in the usual manner: at the first order of potentials, it is given by
\begin{equation}
R = -6\ddot{\Phi} + 2\nabla^{2}(2\Phi - \Psi) \ ,
\label{eq3}
\end{equation}
where a dot denotes the derivative with respect to time and $\nabla$ represents the spatial gradient.
This gives the left-hand side of Eq.~(\ref{eq1}).

Next we consider the right-hand side of Eq.~(\ref{eq1}).
Since the occupation number of the axion in the dark matter halo is huge, we can treat it as a classical scalar field.
The axion field satisfies the Klein-Gordon equation in the flat space-time at the leading order, and the solution is given by the superposition of waves of different frequencies.
Since the typical scale of the dark matter halo, $(10~\text{kpc})^{-1} \sim 10^{5}H_{0}$, is much smaller than the mass of the axion, $m \sim 10^{-23}~\text{eV} \sim 10^{10}H_{0}$, we can assume that the axion field oscillates monochromatically with frequency of its mass.
Under these assumptions, we can write the energy density $\rho$ and pressure $p$ of the axion in the following form:
\begin{equation}
\rho \simeq \rho_{\text{DM}}, \quad p \simeq -\rho_{\text{DM}}\cos(2mt) \ ,
\label{eq4}
\end{equation}
where $\rho_{\text{DM}}$ is a constant.
The typical energy density of the dark matter halo is about $0.3~\text{GeV} / \text{cm}^{3}$.
The negative sign of the pressure is just a convention of choosing a phase.
Therefore, the trace of the energy-momentum tensor of the axion can be written as
\begin{equation}
T = -\rho + 3p \simeq -\rho_{\text{DM}}[1 + 3\cos(2mt)] \ .
\label{eq5}
\end{equation}

From the above results, we can rewrite Eq.~(\ref{eq1}) as follows:
\begin{equation}
-6\ddot{\Phi} + 2\nabla^{2}(2\Phi - \Psi) = \rho_{\text{DM}}[1 + 3\cos(2mt)] \ .
\label{eq6}
\end{equation}
Now let us separate the gravitational potential $\Phi~(\Psi)$ into the time-independent part $\Phi_{0}~(\Psi_{0})$ and the time-dependent part $\delta\Phi~(\delta\Psi)$.
To this aim, we should recall the Poisson equation derived from the time-time component of the Einstein equation
\begin{equation}
2\nabla^{2} \Psi_0 = \rho_{\text{DM}} \ .
\label{eq7}
\end{equation}
We also have the equation $\Psi_0 = \Phi_0$ from the traceless part of the space-space component of the Einstein equation.
Thus, we obtain the equation determining the time dependence of the gravitational potential $\delta\Phi$,
\begin{equation}
-6\delta\ddot{\Phi} = 3\rho_{\text{DM}}\cos(2mt) \ .
\label{eq8}
\end{equation}
The above equation can be easily solved as
\begin{equation}
\delta\Phi = \frac{\pi G\rho_{\text{DM}}}{m^{2}}\cos(2mt) \ ,
\label{eq9}
\end{equation}
where we wrote $8\pi G$ explicitly.
Note that $\delta\Phi \ll \Phi_0$ because $k^2 \ll m^2$ in the present situation~\cite{140212}.

They calculated the timing residuals of the signal from the pulsar and showed that the axion dark matter has the same effect on the pulsar timing measurements as gravitational wave background with characteristic strain
\begin{align}
h_{\text{c}} &= 2\sqrt{3}|\delta\Phi| \nonumber\\
&= 2 \times 10^{-15} \left( \frac{\rho_{\text{DM}}}{0.3~\text{GeV} / \text{cm}^{3}} \right) \left( \frac{10^{-23}~\text{eV}}{m} \right)^{2},
\label{eq10}
\end{align}
at frequency
\begin{equation}
f \equiv \frac{\omega}{2\pi} = 5 \times 10^{-9}~\text{Hz} \left( \frac{m}{10^{-23}~\text{eV}} \right).
\label{eq11}
\end{equation}
This signature is detectable in the planned SKA pulsar timing array experiments.

\section{Axions in $f(R)$ Theory}
In the previous section, we explained how the axion dark matter produces the oscillating gravitational potential in Einstein's theory and the oscillation can be detected through the observation of pulsar timing residuals.
The aim in this paper is to extend the analysis to $f(R)$ theory.
In this section, we will show how to obtain the gravitational potential from axion oscillations in $f(R)$ theory and discuss two specific models.

The action of $f(R)$ theory is given by
\begin{equation}
S = \frac{1}{2} \int d^{4}x\sqrt{-g}[R + f(R)] + S_{\text{m}} \ ,
\label{eq12}
\end{equation}
where $f(R)$ is a function of the Ricci scalar $R$, and $S_{\text{m}}$ is the action for matter fields.
Hereafter, we consider the axion field as the matter.
We assume $f(R) \ll R$ and $f_{R} \equiv f'(R) \ll 1$ so that the deviation from Einstein's theory is small.
The variation of the action with respect to the metric gives the field equation:
\begin{equation}
G_{\mu\nu} - \frac{1}{2}g_{\mu\nu}f + (R_{\mu\nu} + g_{\mu\nu}\Box - \nabla_{\mu}\nabla_{\nu})f_{R} = T_{\mu\nu} \ ,
\label{eq13}
\end{equation}
where $G_{\mu\nu} \equiv R_{\mu\nu} - (1 / 2)g_{\mu\nu}R$ is the Einstein tensor.
The trace of this equation gives
\begin{equation}
-R - 2f + (R + 3\Box)f_{R} = T \ .
\label{eq14}
\end{equation}
We assume that the spatial derivative of $f_{R}$ is much smaller than the time derivative of it, i.e. $\Box f_{R} \simeq -\ddot{f}_{R}$.
Then, we obtain
\begin{equation}
3\ddot{f}_{R} + R = -T \ ,
\label{eq15}
\end{equation}
or equivalently,
\begin{equation}
3f''(R)\ddot{R} + 3f'''(R)\dot{R}^{2} + R = -T \ ,
\label{eq16}
\end{equation}
where we used the approximations $f \ll R$ and $f_{R}\ll 1$.
Since the axion field minimally couples to gravity, we can use the same form for the trace of the energy-momentum tensor of the axion, $T$, as the previous one (\ref{eq5}).
We will use Eq.~(\ref{eq16}) to determine the time-dependent part of the Ricci scalar.

Now, let us consider two specific models.
First, we discuss the $f(R) \propto R^{2}$ model, which can be solved exactly.
Second, we consider the more realistic model known as the Hu-Sawicki model~\cite{070910}.
While it is known that this model can pass both cosmological and solar system tests, we will see that there is a tension between the Hu-Sawicki model and the axion dark matter for some parameters.

\subsection{$f(R) = R^{2} / 6M^{2}$ model}
Let us consider a simple model given by
\begin{equation}
f(R) = \frac{R^{2}}{6M^{2}} \ ,
\label{eq17}
\end{equation}
where $M$ is a constant mass scale.
When we discuss the model in terms of the scalar-tensor theory, $M$ is indeed the mass of the scalar field.
This type of model was introduced by Starobinsky to explain the inflationary universe~\cite{800524}.
Now, however, we use this model in the context of the dark energy.

Now, the field equation (\ref{eq16}) becomes
\begin{equation}
\frac{1}{M^{2}}\ddot{R} + R = \rho_{\text{DM}}[1 + 3\cos(2mt)] \ ,
\label{eq18}
\end{equation}
and the solution is given by
\begin{equation}
R = \rho_{\text{DM}} + \frac{3\rho_{\text{DM}}}{1 - (2m / M)^{2}}\cos(2mt) \ .
\label{eq19}
\end{equation}
Here, we ignored the homogeneous solutions which oscillate freely with frequency $M$, and we will discuss this point at the end of this subsection.

\begin{figure}
\includegraphics[width = 150pt]{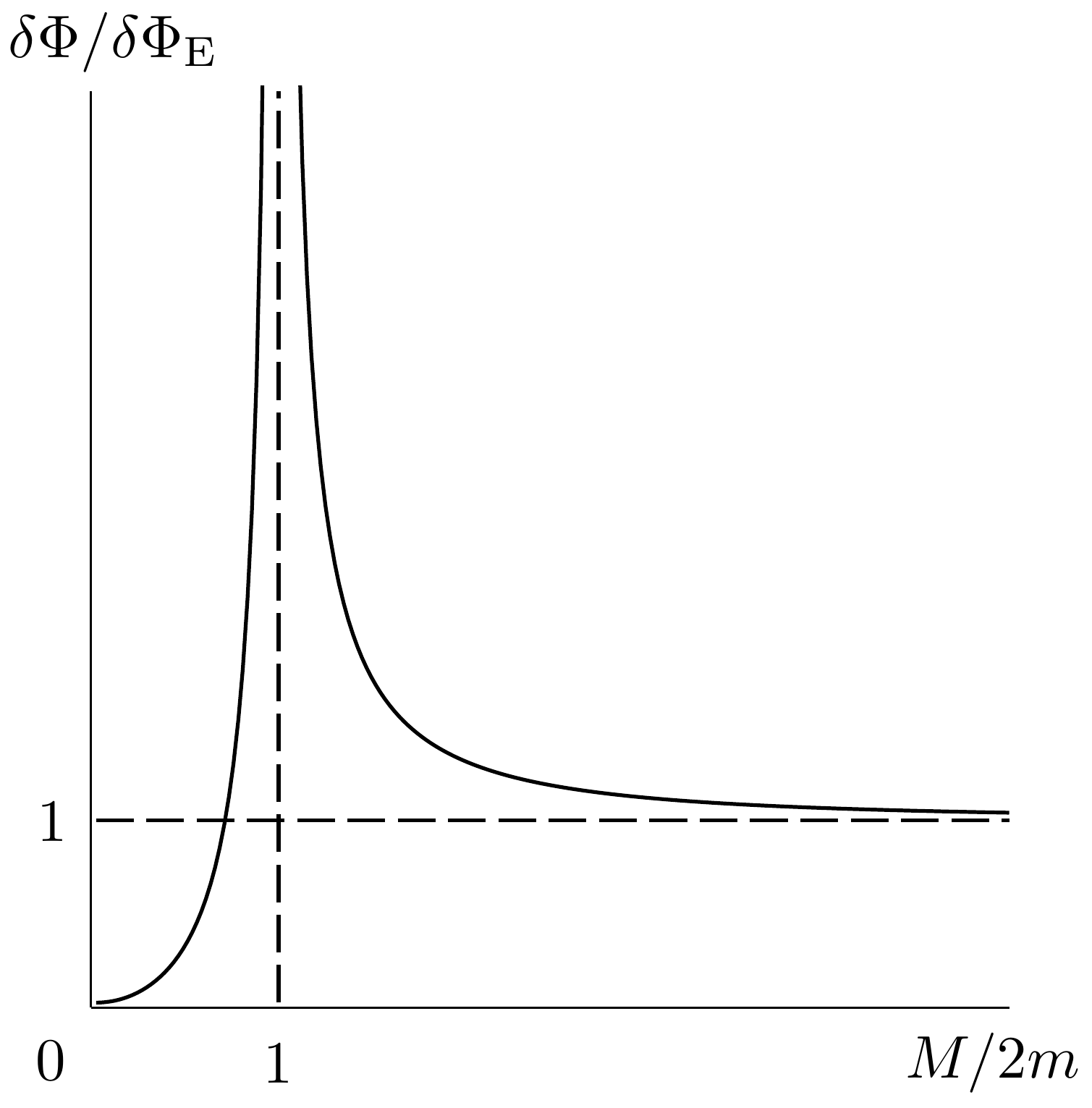}
\caption{The amplitude of the gravitational potential in $R^{2}$ model normalized by the value in Einstein's theory.
Note that we plotted the absolute value because the sign is not important.}
\label{Fig1}
\end{figure}

Following the same procedure done in the case of Einstein's theory, we obtain the time-dependent part of the gravitational potential as follows:
\begin{equation}
\delta\Phi = \frac{1}{1 - (2m / M)^{2}}\frac{\pi G\rho_{\text{DM}}}{m^{2}}\cos(2mt) \ .
\label{eq20}
\end{equation}
Therefore, in the $R^{2}$ model, we obtained the amplitude of the gravitational potential relative to that in Einstein's theory:
\begin{equation}
\frac{\delta\Phi}{\delta\Phi_{\text{E}}} = \frac{1}{1 - (2m / M)^{2}} \ ,
\label{eq21}
\end{equation}
where $\delta\Phi_{\text{E}}$ is the amplitude of the gravitational potential predicted in Einstein's theory.
This result is illustrated in Fig.~\ref{Fig1}.
In the large $M$ limit, $M \gg 2m$, the prediction in $f(R)$ theory is the same as in Einstein's theory.
This can be understood from the form of $f(R) = R^{2} / 6M^{2}$.
In fact, when the mass of the scalar field $M$ becomes large, $f(R)$ can be neglected compared to the Ricci scalar $R$.
Thus, Einstein's theory is reproduced in this limit.
In the opposite limit, $M \ll 2m$, the amplitude of the gravitational potential goes to zero.
Hence, in this case, it would be difficult to detect the oscillation of the gravitational potential.
When the mass scale $M$ gets close to the frequency of the pressure, $2m$, resonance would occur and the amplitude of the gravitational potential would be dramatically amplified.
Of course, the amplitude cannot reach to infinity: the approximation becomes worse when the oscillating part of $f(R)$ cannot be ignored compared to the Ricci scalar, $R$.

Now, we make a comment on homogeneous solutions ignored before.
It is pointed out by Starobinsky that the homogeneous solutions decay in the expansion universe and can be completely ignored at the present time~\cite{071000}.
In addition, it is supposed that such scalar degrees of freedom should be highly suppressed by some mechanisms in the solar system scale in order not to mediate the so-called fifth force.
For example, taking into account the interactions, the stabilization mechanism called chameleon mechanism~\cite{040227} or Vainshtein mechanism~\cite{720501} would work and such degrees of freedom might be killed in the solar system scale.
However, if such modes were alive in the dark matter halo scale for some reasons and the mass scale $M$ were sufficiently close to the frequency of the pressure, $2m$, a beat would occur with a frequency $|M - 2m|$.
In this situation, after averaging over the time scale corresponding to the high-frequency $M \sim 2m$, we would observe the beat frequency, $|M - 2m|$.
If such a thing happened, the detectable mass range of the axion by the pulsar timing observation would shift to more heavy mass regions.

\subsection{Hu-Sawicki model}
In the previous subsection, we discussed the simplest $f(R)$ model which can be solved exactly.
Now, in this subsection, we consider a more realistic model.

While there are several $f(R)$ models that explain the late time acceleration of the universe and also pass the solar system tests, now let us focus on the specific model known as the Hu-Sawicki model:
\begin{equation}
f(R) = -\mu R_{\text{c}}\frac{(R / R_{\text{c}})^{2n}}{(R / R_{\text{c}})^{2n} + 1} \ ,
\label{eq22}
\end{equation}
where $n, \mu, R_{\text{c}} > 0$.
For this model to mimic the $\Lambda$CDM model, $\mu R_{\text{c}} \simeq 2\Lambda$ is needed, where $\Lambda$ is the cosmological constant.
Since the energy density of the dark matter halo is much larger than the cosmological critical density, we can assume $R \gg R_{\text{c}}$.
In this limit, Eq.~(\ref{eq22}) takes the following form:
\begin{equation}
f(R) \simeq -\mu R_{\text{c}} \left[ 1 - (R / R_{\text{c}})^{-2n} \right].
\label{eq23}
\end{equation}
Note that the Starobinsky model~\cite{071000} has the same form as Eq.~(\ref{eq23}) in the high curvature limit.

In order to pass the local gravity tests, the Ricci scalar should oscillate around its average value $R_{0} = \rho_{\text{DM}}$.
The mass scale of this model is given by
\begin{equation}
M^{2} \equiv \frac{1}{3f''(R_{0})} \simeq \frac{R_{\text{c}}}{6n(2n + 1)\mu} \left( \frac{\rho_{\text{DM}}}{R_{\text{c}}} \right) ^{2n + 2}.
\label{eq24}
\end{equation}
Using $R_{\text{c}} \simeq 2\Lambda / \mu$ and plausible cosmological parameters~\cite{150209}, the mass is roughly evaluated as
\begin{equation}
M \sim 1.5\mu \times 10^{-23}~\text{eV} \ ,
\label{eq25}
\end{equation}
for $n = 1$.
This rough estimate tells us that $M$ has a value around the critical mass, $2m$, for $\mu = \mathcal{O}(1)$.
Since $M$ is strongly dependent on $n$ [see Eq.~(\ref{eq24})], $M$ can be larger or smaller compared to $2m$.

When $M \ll 2m$, completely the same situation as $R^{2}$ model is realized and the amplitude of the gravitational potential is given by Eq.~(\ref{eq20}).
This is because the amplitude of the Ricci scalar is much smaller than its average value in this limit and the field equation (\ref{eq16}) is reduced to Eq.~(\ref{eq18}).
Hence, this behavior should be universal for more general models which pass the local gravity tests.

When $M \gtrsim 2m$, however, a problem arises.
Since $f'''(R) / f''(R) \sim 1 / R$, we can evaluate
\begin{equation}
\frac{f'''(R)\dot{R}^{2}}{f''(R)\ddot{R}} \sim \frac{\delta\dot{R}^{2}}{R\delta\ddot{R}} \sim \frac{\delta R}{R} \ ,
\label{eq26}
\end{equation}
where $\delta R$ is the oscillating part of $R$.
Therefore, once $\delta R$ becomes of the order of $R_{0}$, the second term of the field equation (\ref{eq16}) prohibits $\delta R$ from oscillating stably.
With numerical calculations, we verified the Ricci scalar diverges for these parameters.
This is also true for the Einstein limit, $M \gg 2m$.
Thus, the Hu-Sawicki model is not compatible with the axion dark matter for these parameters.
Note that other viable $f(R)$ models also suffer from the same problem.

\begin{figure}
\includegraphics[width = 200pt]{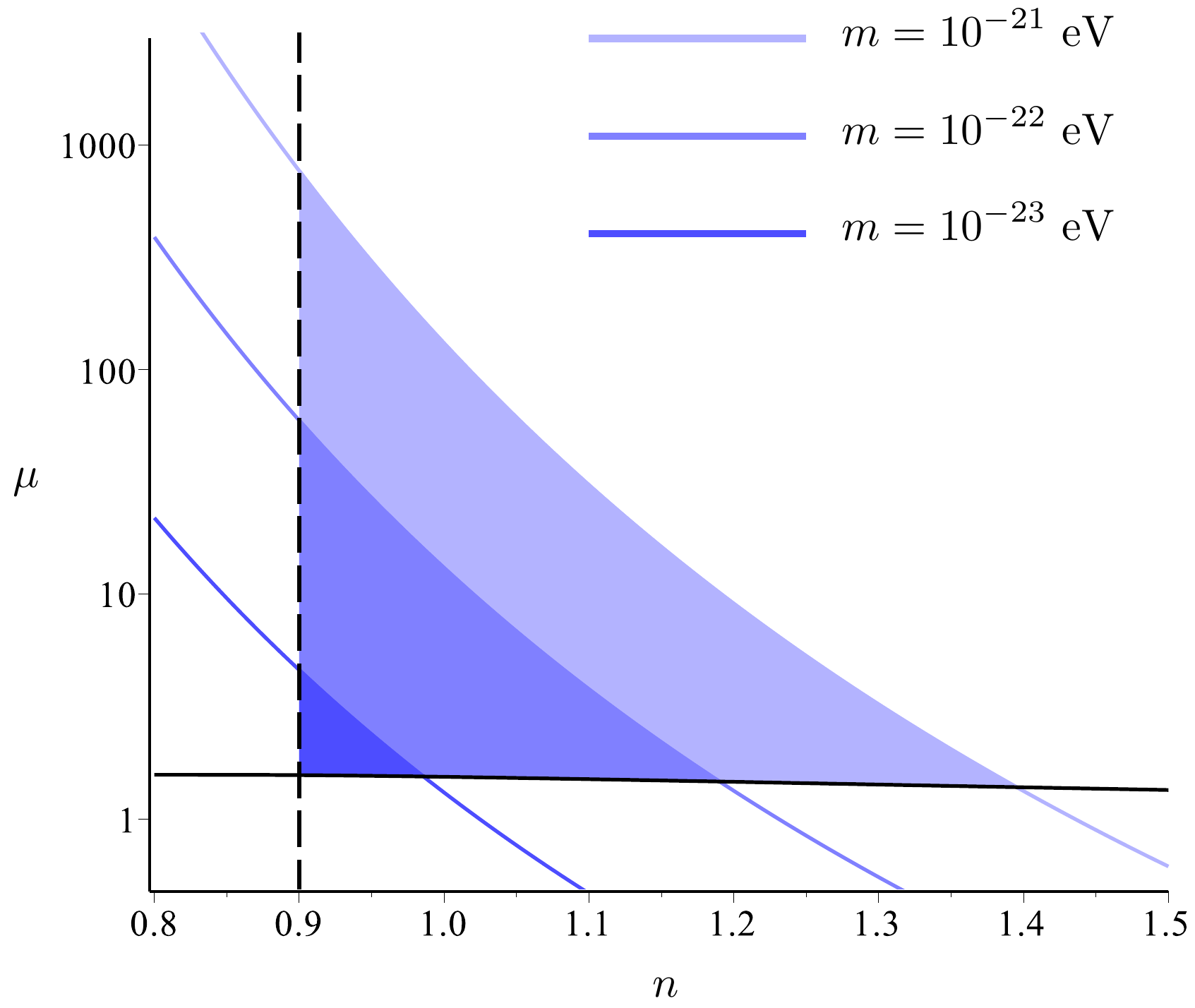}
\caption{
The constraints on the Hu-Sawicki model.
The axion dark matter model and the Hu-Sawicki model are compatible in the shaded regions.
}
\label{Fig2}
\end{figure}

In order to avoid the instability of oscillations, the condition $M \lesssim 2m$ is needed.
This constraint is shown in Fig.~\ref{Fig2} with other constraints from cosmological and local gravity tests~\cite{100623}:
The three downward-sloping curves are the upper bounds on $\mu$ for three different axion masses.
The almost horizontal line denotes the lower bound on $\mu$ from cosmological tests.
The vertical line corresponding to $n = 0.9$ represents the lower bound on $n$ from the local tests.
From Fig.~\ref{Fig2}, the ultralight axion dark matter and the Hu-Sawicki model are compatible only in the certain parameter regions~(shaded in Fig.~\ref{Fig2}).
Note that the upper bounds on $\mu$ are somewhat underestimated:
From numerical calculations, we found that the upper bounds on $\mu$ are about 5 times smaller than the roughly estimated values illustrated in Fig.~\ref{Fig2}.
Of course, since the Hu-Sawicki model works well on large scales, it might be natural to modify the Hu-Sawicki model on small scales to circumvent this instability problem.

If the axion dark matter were detected by pulsar timing experiments, we can determine the axion mass $m$ from the oscillation frequency
and the mass scale $M$ of $f(R)$ model from  the amplitude of oscillation Eq.~(\ref{eq20}).
In the Hu-Sawicki model , since $M$ monotonically increases as $\mu$ and $n$ increase, it has the minimum value, $M_{\text{min}}$, 
corresponding to the lower bounds for $\mu$ and $n$.
Numerically, we obtain the minimum value as
\begin{equation}
M_{\text{min}} \sim 0.76 \times 10^{-23}~\text{eV} \ .
\label{eq27}
\end{equation}
Hence, if the observed mass $M$ were lower than Eq.~(\ref{eq27}), the Hu-Sawicki model would be excluded.

\section{Conclusion}
We studied the pulsar timing signal from the ultralight axion field in $f(R)$ theory.
First, we discussed the simplest $f(R) = R^{2} / 6M^{2}$ model.
Then, it turned out that the amplitude of the gravitational potential in this model is enhanced or suppressed depending on the mass parameter $M$ compared to the case in Einstein's theory.
If $M$ is larger than the frequency of the pressure, $2m$, the results in Einstein's theory are reproduced.
On the other hand, if $M$ is smaller than $2m$, the amplitude is suppressed and difficult to be detected.
Furthermore, when $M$ approaches $2m$, the amplitude is dramatically amplified due to the resonance.

Next we discussed the Hu-Sawicki model.
Although the Hu-Sawicki model is known to pass both cosmological and solar system tests, we showed that this model is not compatible with the ultralight axion dark matter for some parameters.
When the mass scale $M$ given by Eq.~(\ref{eq24}) is much smaller than $2m$, completely the same situation as $R^{2}$ model is realized.
In this case, unfortunately, the amplitude is too small to be detected by near-future experiments.
Meanwhile, when $M$ reaches  $2m$, the oscillation cannot be stable owing to  the ``nonlinear" term of the field equation.
Remarkably, the model does not work even in the Einstein limit, $M \gg 2m$, for the same reason.
This gives rise to the new constraint on the Hu-Sawicki model.
In order to circumvent this instability problem, a modification on small scales would be needed.
In fact, if the detected mass scale $M$ were lower than $M_{\text{min}} \sim 0.76 \times 10^{-23}~\text{eV}$, the Hu-Sawicki model would be excluded.

\acknowledgements
This work was supported by JSPS KAKENHI Grant No. 25400251, MEXT KAKENHI Grants No. 26104708 and No. 15H05895.

\bibliographystyle{apsrev4-1}
\bibliography{Paper}

\end{document}